\documentclass[final,5p,times,onecolumn]{elsarticle}

\usepackage{graphicx,color,soul}
\usepackage{amsmath,amssymb,bm}

\def\Xint#1{\mathchoice
   {\XXint\displaystyle\textstyle{#1}}%
   {\XXint\textstyle\scriptstyle{#1}}%
   {\XXint\scriptstyle\scriptscriptstyle{#1}}%
   {\XXint\scriptscriptstyle\scriptscriptstyle{#1}}%
   \!\int}
\def\XXint#1#2#3{{\setbox0=\hbox{$#1{#2#3}{\int}$}
     \vcenter{\hbox{$#2#3$}}\kern-.5\wd0}}

\def\dashint{\Xint-}

\usepackage{epsfig}
\usepackage{amsmath}
\usepackage{graphicx}
\usepackage{mathptmx}      
\usepackage{latexsym}
\usepackage{bm}
\usepackage{comment}

\journal{Physics Letters B}

\usepackage{supertabular} 
\usepackage{placeins}
\usepackage{epsfig}
\usepackage{graphicx}

\usepackage{epstopdf}

\begin{document} 
\begin{frontmatter}

\title{Counting states and the Hadron Resonance Gas: Does X(3872) count ? }

\author{Pablo G. Ortega}
\ead{pgortega@usal.es}
\author{David R. Entem}
\ead{entem@usal.es}
\author{Francisco Fern\'andez}
\ead{fdz@usal.es}
\address{Grupo de F\'isica Nuclear and Instituto Universitario de F\'isica 
Fundamental y Matem\'aticas (IUFFyM), Universidad de Salamanca, E-37008 
Salamanca, Spain}
\author{Enrique Ruiz Arriola}
\ead{earriola@ugr.es}
 \address{Departamento de
  F\'{\i}sica At\'omica, Molecular y Nuclear  and Instituto Carlos I
  de F{\'\i}sica Te\'orica y Computacional \\ Universidad de Granada,
  E-18071 Granada, Spain.} 

\date{\today}

\begin{abstract}
We analyze  how the renowned X(3872), a weakly bound state right
below the $D \bar D^*$ threshold, should effectively be included
 in a hadronic representation of the QCD partition function. This
can be decided by analyzing the $D \bar D^*$ scattering phase-shifts
in the $J^{PC}=1^{++}$ channel and their contribution to the level
density in the continuum from which the abundance in a hot medium can
be determined. We show that in a purely molecular picture the bound
state contribution cancels the continuum providing a vanishing
occupation number density at finite temperature and the $X(3872)$ does
not count below the Quark-Gluon Plasma crossover happening at $T \sim
150$MeV. In contrast, within a coupled-channels approach, for a
non vanishing $c \bar c$ content the cancellation does not occur due
to the onset of the $X(3940)$ which effectively counts as an
elementary particle for temperatures above $T \gtrsim 250$MeV. Thus, a
direct inclusion of the $X(3872)$ in the Hadron Resonance Gas is not
justified. We also estimate the role of this cancellation in
  X(3872) production in heavy-ion collision experiments in terms of
  the corresponding $p_T$ distribution due to a finite energy resolution.
\end{abstract}



\end{frontmatter}


\section{Introduction}

Counting hadronic states below a certain mass and QCD thermodynamics
at finite temperature in a box with a finite volume are intimately
related. However, while the counting process requires an individual
knowledge of the mass spectrum, thermodynamics generally implies a
collective information. Experimentally both pieces of information are
obtained by different means; while the single states are determined
one by one by spectroscopic measurements and the analysis of hadronic
reactions the determination of thermal properties acquires a more
macroscopic nature such as in ultra-relativistic heavy ions
collisions. Within such context basic objects are occupation
  numbers and their corresponding transverse momentum and rapidity
  distributions which are extracted from experiment if assumptions on
  the fireball freeze-out dynamics are implemented~(see
  e.g. \cite{florkowski2010phenomenology} and references therein.).

Specifically, the coupling of {\it any} hadronic state to
a heat bath at temperature $T$ is universally given by the Boltzmann
factor,
\begin{eqnarray}
  Z= \sum_n e^{-M_n/T} = \int dM \rho(M) e^{-M/T} \, . 
\end{eqnarray}
Here $M_n$ mean the QCD (discretized) eigenstates in a finite box
which due to confinement are colour neutral and $\rho(M)=N'(M)$ is the
density of states where $N(M)$ is the cumulative number of states
\begin{eqnarray}
  N(M) = \sum_n \theta (M-M_n)  \, . \label{eq:Ncum} 
\end{eqnarray}
At small temperatures we and due to confinement we expect hadronic
states to saturate the partition function. Based on the quantum virial
expansion in quantum mechanics~\cite{Beth:1937zz} and quantum field
theory \cite{Dashen:1969ep} a genuine hadronic representation was
derived in terms of the S-matrix in the continuum limit, $N(M) = {\rm
  Tr} \log S / 2\pi i$ where the cumulative number becomes a real,
non-integer, number. In this case, the actual implementation of this
approach requires, besides taking the box volume to infinity,
consideration of interactions among multiparticle states built from
the asymptotic scattering free states.  This means that only ground
states of the strong interaction (in the confined phase) should be
used in constructing the Fock space. At sufficiently low temperatures,
lowest masses dominate and one has to successively incorporate $\pi$,$2
\pi$, $3 \pi$, $\eta$, $K$, etc. While two-body states can be
described by phase-shifts~\cite{Beth:1937zz}, the three body
contribution is a complex problem, making 
 the approach unmanageable without further approximations 
 ( See Ref.~\cite{Lo:2017sde} for a recent and promising attempt to address 
 the $(N>2)$-body problem in a model-independent way). 
Fortunately, as pointed out soon after
\cite{Dashen:1969ep} the role of narrow
resonances~\cite{Dashen:1974jw} and effective
elementarity~\cite{Dashen:1974yy} was shown to reduce the
thermodynamics of QCD in the confined phase to a Hadron Resonance Gas
(HRG), where the hadronic states are identified and counted {\it one
  by one} effectively entering the partition function as single
particle states~\footnote{This way one handles, e.g., three body
  interactions as two-step processes mediated by resonant scattering;
  if $2 \pi \to \rho$, then $3 \pi \to \pi \rho \to \omega, A_1$ and
  so on.}. In the mid 60's Hagedorn analyzed the mass-level density
$\rho(M)= N'(M)$ and, conjecturing the validity of the HRG, {\it
  predicted} the bulk of states at higher masses, which later on were
experimentally confirmed~\cite{Hagedorn:1965st}. The more recent
updates in~\cite{Broniowski:2000bj,Broniowski:2004yh} proposed to use
directly $N(M)$ as the relevant quantity, which features explicitly the
notion of counting as shown in Eq.~(\ref{eq:Ncum}). Overall, resonance
widths (in the Breit-Wigner approximation) have the effect of
reshuffling the mass distribution around the resonance mass value and
hence increasing, regularizing, i.e. making it smooth, and
``de-quantizing'' this
quantity~\cite{Arriola:2012vk,Broniowski:2016hvt}. 

The commonly accepted reference for hadronic states is the Particle
Data Group (PDG) table~\cite{Patrignani:2016xqp}, a compilation reflecting
a consensus in the particle physics community whose cumulative number
$N_{\rm PDG}(M)$ has most spectacularly been checked by the
computation of the trace anomaly, $\epsilon- 3 P = T^5 \partial_T
(\log Z /T^3)/V$, on the
lattice~\cite{Borsanyi:2013bia,Bazavov:2014pvz,Borsanyi:2016ksw} at
temperatures $T \lesssim 200 {\rm MeV}$ below the crossover to the
Quark-Gluon Plasma (QGP) phase. It is worth noting that this agreement
between the WB~\cite{Borsanyi:2013bia,Borsanyi:2016ksw} and the
HotQCD~\cite{Bazavov:2014pvz} lattice collaborations and with the HRG
has come after many years of frustration and controversy. Width
effects reflect the mass reshuffling by increasing the trace anomaly
and agree still within the lattice
uncertainties~\cite{Arriola:2012vk,Broniowski:2016hvt} (see
e.g. Ref.~\cite{Arriola:2014bfa} for a pedagogical exposition and
overview).

These results suggest that {\it all} states listed by the PDG should
also be counted in the cumulative number as genuine contributions to
the QCD partition function and hence directly included in the
HRG. However, in a remarkable and forgotten paper Dashen and Kane
pointed out the possibility that not all hadron states should be
counted on a hadronic scale~\cite{Dashen:1974ns} as they become
fluctuations in a mass-spectrum coarse grained sense. The deuteron, a
$J^{PC}=1^{++}$ np composite, was prompted as a non-controversial
example where the weak binding effect is compensated by the nearby np
continuum yielding an overall vanishing contribution. The
basic idea was that certain interactions do not generate new states
but simply reorder the already existing ones (see
\cite{Arriola:2015gra} for an explicit figure of the cumulative number
in the deuteron channel).

The possibility of having loosely bound states near the charm
threshold, i.e.  Charm Molecules, was envisaged long
ago~\cite{Nussinov:1976fg}.  Actually, the discovery of the state
$X(3872)$ in 2003 by the Belle Collaboration in the exclusive
$B^\pm\to K^\pm \pi^+ \pi^- J/\psi$ decay ~\cite{Choi:2003ue} has
initiated a new era in hadronic spectroscopy. This state decays
through the $J/\psi\rho$ and $J/\psi\omega$ channels which are
forbidden for a $c\bar c$ configuration and has $J^{PC}=1^{++}$ as
concluded by the LHCb Experiment by means of the five-dimensional
angular analysis of the process $B^+\rightarrow K^+X(3872)$ with
$X(3872)\rightarrow J/\psi \rho^0\rightarrow J/\psi
\pi^+\pi^-$~\cite{PhysRevD.92.011102}. As a natural consequence
  this state has entered the PDG with a current binding energy of
  $B_X \equiv M_X-M_{D^0}-M_{\bar D^{0*}}=0.01(18)$MeV~\cite{Patrignani:2016xqp}.

The proliferation of new X,Y,Z states (see~\cite{Lebed:2016hpi} for a
recent review) and their inclusion in the PDG poses the natural
question whether or not these states have some degree of redundancy in
order to build the hadron spectrum. The possibility that this might
happen for some weakly bound X,Y,Z states has been suggested
recently~\cite{Arriola:2014bfa,Arriola:2015gra}. In the present paper
we analyze this issue for the renowned $X(3872)$ case by analyzing for
the first time $D \bar D^*$ scattering and show that the answer to
this question depends on the particular dynamics of the system. 
  This is particularly relevant as recently the $p_T$ distribution of
  the $X(3872)$ in pp collisions have been determined both
  theoretically~\cite{Artoisenet:2009wk,Meng:2013gga} and
  experimentally by CMS~\cite{Chatrchyan:2013cld} and
  ATLAS~\cite{Aaboud:2016vzw} and the possible implications on the
  molecular content have been examined~\cite{Esposito:2015fsa}.  Our
  results apply specifically to $X(3872)$ production in heavy-ion
  collisions, for which no experiments exist yet.

\section{Counting states and their abundance}

For an elementary and free state with $g$-degrees of freedom and mass $m$ 
in a medium with
temperature $T$ the average density of particles is given by 
\begin{equation}\label{ec:Nocup1}\begin{split}
\bar n= \frac{\langle N \rangle_T}{V} &=
\int \frac{d^3 k}{(2\pi)^3} \frac{g}{e^{\sqrt{k^2+m^2} /T} + \eta}  \\ 
&= \frac{T^3}{2\pi^2} \sum_{n=1}^\infty g \frac{(-\eta)^{n+1}}{n}
  \left( \frac{m}{T} \right)^2 K_2(n m/T) \, , 
\end{split}\end{equation}
where $K_2(x)$ is the modified Bessel function and $\eta=\mp 1$ for
bosons/fermions respectively~\footnote{In practice the Boltzmann
  approximation (i.e., just keeping $n=1$) is sufficient for low
  temperatures}. In the case of composite particles or two-body
interacting particles, according to the quantum virial
expansion~\cite{Beth:1937zz,Dashen:1969ep} the effects of interactions
can be expressed in terms of scattering phase shifts
\begin{eqnarray}
n(T) = \int \frac{d^3 p}{(2\pi)^3} dm \frac{g}{e^{\sqrt{p^2+m^2} /T} + \eta} \rho(m) \, , 
\end{eqnarray}
where  
\begin{eqnarray}\label{ec:stateden}
\rho(m) = \frac1{\pi} \frac{d \delta}{d m} \, . 
\end{eqnarray}

For a narrow resonance with mass $m_R$ and width $\Gamma_R \to 0$ the
phase-shift can be described by a Breit-Wigner shape $\delta(m)=
\tan^{-1}[(m-m_R)/\Gamma_R]$ so that $\delta'(m) \to \pi \delta
(m-m_R)$, and their contribution becomes that of an elementary
particle with mass $m_R$~\cite{Dashen:1974jw}. For instance, in the
case of $\pi\pi$ scattering in the isovector channel the contribution
is given by the corresponding $\rho$ resonance. Interestingly,
cancellations among different $\pi\pi$ and $\pi K$ channels have been
reported~\cite{Venugopalan:1992hy,GomezNicola:2012uc,Broniowski:2015oha,Friman:2015zua}
implying, for instance, that the lowest $0^{++}$ isoscalar state,
quoted as the $f_0(500)$ in the PDG and also known as the $\sigma$
meson cancels the isotensor contribution, i.e. 
$\delta_{00}+5 \delta_{0,2} =0 $ within uncertainties for $\sqrt{s} \le 900$MeV~\cite{GarciaMartin:2011cn}. This is essentially a
cancellation between the attraction in the $I=0$ channel generating
the resonance and a repulsive in the $I=2$ channel possibly
triggered by the finite pion-size generating a hard core.


Here, we address a different type of cancellation unveiled by Dashen
and Kane~\cite{Dashen:1974ns}, namely the fact that for a certain type
of loosely bound state, the contribution may effectively vanish. For
completeness, let us review briefly their argument.  The cumulative
number in a given channel in the continuum with threshold $M_{\rm th}$
is
\begin{eqnarray}
N(M)= \sum_n \theta(M-M_n^B)
  + \frac1\pi \sum_{\alpha=1}^{K} [\delta_\alpha (M)-\delta_\alpha (M_{\rm th})] \, .
\label{eq:ncum} 
\end{eqnarray}
Here the bound states masses $M_n^B$ have been explicitly separated from
scattering states written in terms of the eigenvalues of the S-matrix,
i.e. $S = U {\rm Diag} (\delta_1, \dots, \delta_K ) U^\dagger$ with
$U$ a unitary transformation for K-coupled channels. With this
definition we have $N(0)=0$, and in the single channel case, in the
limit of high masses $M \to\infty$ becomes
\begin{eqnarray}
N(\infty)=n_B +
\frac1{\pi} [\delta(\infty)-\delta(M_{\rm th})]=0
\end{eqnarray}
due to Levinson's theorem which is the statement that the total number
of states does not depend on the interaction. In the NN channel where
$M_{\rm th}= 2M_N$ the appearance of the deuteron changes rapidly at
$M= 2 M_N -B_d $ by one unit so that $N ( 2M_N-B_d +0^+)- N ( 2M_N-B_d
-0^+)=1 $, but when we increase the energy this number decreases
slowly to zero at about pion production threshold $N ( 2M_N+ m_\pi)- N
( 2M_N-B_d -0^+) \sim 0 $. This features are depicted in
Ref.~\cite{Arriola:2015gra} for $\sqrt{s}$ up to 3.5GeV.  A direct
consequence of this is that the deuteron abundance at hadronic
temperatures will be almost zero! This effect is explicitly seen in
the np virial coefficient at rather low
temperatures~\cite{Horowitz:2005nd}.

\section{The X(3872) and $D \bar D^*$ Scattering in the molecular picture}

While $X(3872)$ is most naturally defined as a pole of the $D\bar D^*$
scattering amplitude, to our knowledge the physically meaningful
phase-shifts have never been explicitly analyzed.  Actually, the QCD
evidence for $X(3872)$ on the lattice has been pointed out
\cite{Prelovsek:2013cra} by analyzing the energy shifts on a finite
volume by means of the L\"uscher's formula where the connection to $D\bar
D^*$ scattering is established.

\begin{figure}[h]
\begin{center}
\epsfig{figure=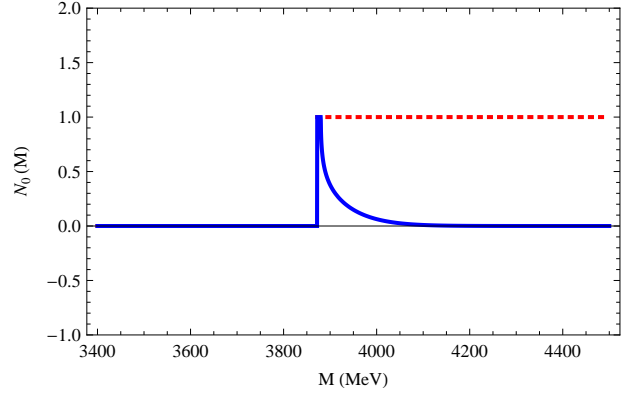,width=0.9\linewidth} 
\end{center}
\caption{Color online: Cumulative number in the $1^{++}$ 
channel as a function of the $D \bar D^*$ mass (in MeV) 
for the $X(3872)$ {\it only} (dotted,red) and the 
full contribution including the continuum (full, blue).}
\label{fig:ncumEFT}
\end{figure}

The weak binding of the $X(3872)$ has suggested in the early studies a
purely molecular nature. It is instructive to analyze scattering
within a purely hadronic picture of contact
interaction~\cite{Gamermann:2009uq}, with the hope that short distance
details can be safely ignored~\footnote{Isospin effects have been
  considered in~\cite{Gamermann:2009uq} where the coupling of the X to
  the neutral and charged components is very similar. Here we will
  ignore the effect and take an average value for the binding.}. If we
take an interaction of the form $V_0 (k',k) = C_0 g(k') g(k) $, the
phase shift is given by (see e.g. Ref.~\cite{Arriola:2013era}),
\begin{eqnarray}
 p \cot \delta_0 (p) &=& - \frac{1}{V_0 (p,p)} \left[1- \frac{2}{\pi}
   \dashint_0^\infty dq \frac{q^2}{p^2-q^2} V_0(q,q) \right] \nonumber
 \\ &=& -\frac1{\alpha_0} + \frac12 r_0 p^2 + \dots 
\label{eq:ERE}
\end{eqnarray}
where in the last line a low momentum Effective Range Expansion (ERE)
has been carried out, identifying $\alpha_0$ with the scattering
length and $r_0$ with the effective range. Fixing $\alpha_0 =
3.14\,{\rm fm}$ and $r_0=1.25\, {\rm fm}$ (see next Section) we get
the phase shift and using Eq.~(\ref{eq:ncum}) we get the cumulative
number including the continuum 
states depicted in Fig.~\ref{fig:ncumEFT} compared with the  case where 
{\it only} the X(3872) is considered~\footnote{We use the
  Gaussian regulator $g(k)=e^{-k^2/\Lambda^2}$ and obtain $C_0 = -1.99
  {\rm fm}$ and $\Lambda= 2.05 {\rm fm}^{-1}$. The pole in the
  scattering amplitude is at $k_X= i 0.43 {\rm fm}^{-1}$ corresponding
  to $M_X=3868$MeV. Note that we disregard isospin effects, see
  Ref.~\cite{Gamermann:2009uq} otherwise.  Other smooth regulators
  give similar results. }. 
This illustrates the point made by Dashen
and Kane~\cite{Dashen:1974ns} in the case of the $X(3872)$, showing
that in the molecular picture the state {\it does not} count in the $D
\bar D^*$ continuum on coarse mass scales of about $\Delta M_{D \bar
  D^*} \sim 200$MeV~\footnote{The resemblance with the deuteron case
  is striking, see Ref.~\cite{Arriola:2015gra} for $\sqrt{s}$ up to
  3.5GeV, where mass scales are about a half, $M_N \sim M_D/2$ and
  $M_d \sim M_X/2$ as in the $X(3872)$. So, the coarse mass scale here
  is $\Delta M_{NN} \sim \Delta M_{D \bar D^*} /2 \sim 100$MeV.}.

\section{The X(3872) and $D \bar D^*$ Scattering in the cluster quark model picture}

The multichannel scattering problem with confined intermediate states
was initiated after the first charmonium evidences based on the
decomposition of the Hilbert space as ${\cal H} = {\cal H}_{c \bar c}
\oplus {\cal H}_{D \bar D}$~\cite{Dashen:1976cf,Eichten:1978tg}. 
  In the multichannel case with permanently confined channels,
  Levinson's theorem is modified \cite{Dashen:1976cf} by subtracting
  the number of bound states of the purely confining potential, $n_c$, so
  that $N(\infty)= n_c$ in Eq.~(\ref{eq:ncum}).

\begin{figure*}[htbp]
\begin{center}
\epsfig{figure=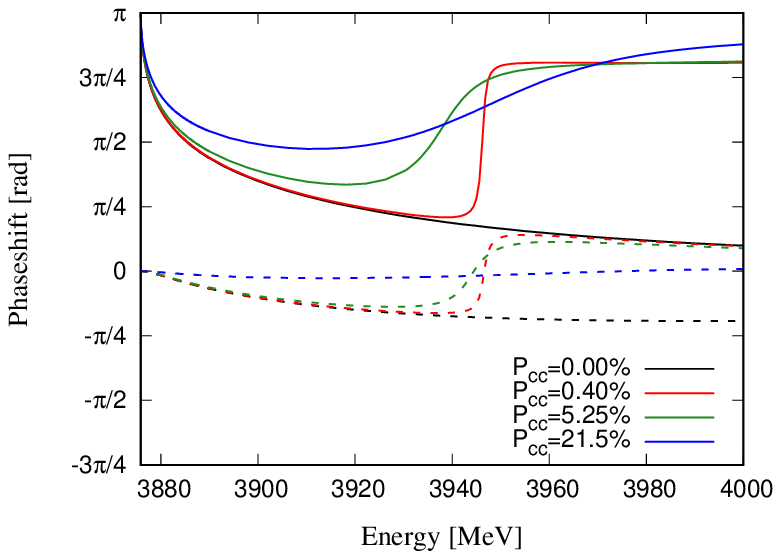,width=0.45\linewidth} 
\epsfig{figure=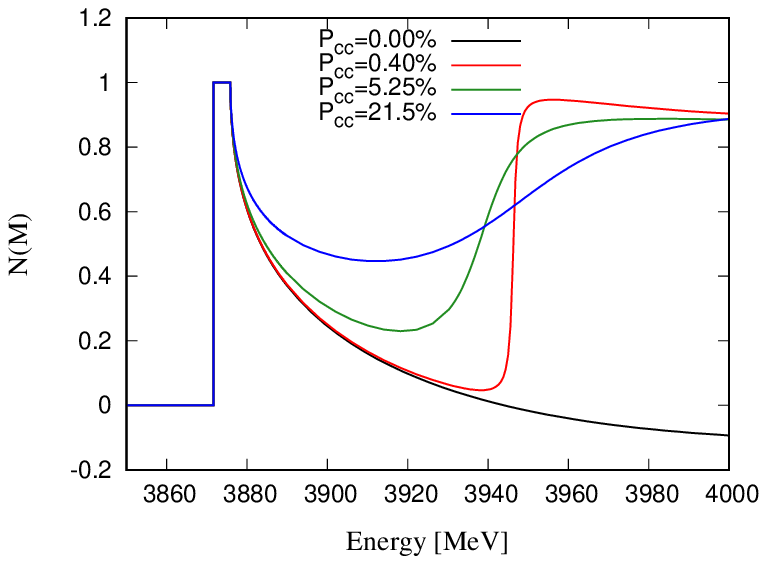,width=0.45\linewidth} 
\end{center}
\caption{Color online: Left panel: S- (solid) and D-wave (dashed)
  phase-shifts in radians as a function of the $D \bar D^*$ invariant
  mass. Right panel: Cumulative number in the $X(3872)$ channel as a
  function of the $D \bar D^*$ mass.}
\label{fig:phases}
\end{figure*}

A coupled-channels calculation which included such decomposition was
addressed in Ref.~\cite{Ortega:2009hj}, performed in the framework of
the constituent quark model (CQM) proposed in
Ref.~\cite{Vijande:2004he}.  This CQM has been extensively used to
describe the hadron phenomenology both in the
light~\cite{PhysRevC.64.058201} and the heavy quark
sectors~\cite{Segovia:2008zz,Segovia:2016xqb}.  In
Ref.~\cite{Ortega:2009hj}, the $X(3872)$ resonance together with the
$X(3940)$ have been explained as two $J^{PC}=1^{++}$ states, being the
$X(3872)$ basically a $D\bar D^\ast+h.c.$ molecule with a small amount
of $2^3P_1$ $c\bar c$ state while the $X(3940)$ is a mixture with more
than $60\%$ of $c\bar c$ structure.  Actually, in the absence of
  mixing, $X(3940)$ becomes a pure $c\bar c$ state, and the only
  confined state in the $J^{PC}=1^{++}$ channel. The aim of  Ref.~\cite{Ortega:2009hj} (extended in Ref.~\cite{Ortega:2012rs}) was to study the
$J^{PC}=1^{++}$ sector including the effect of the closest $c\bar c$
states in the dynamics of the $D\bar D^\ast$ channel.  For simplicity,
we will consider the $D^{(\ast)}$ mesons as effectively stable, due to
their narrow width, and we will only consider the isospin-zero $D\bar
D^\ast$ channel, as the $c\bar c-DD^\ast$ coupling mechanism occurs
solely in $I=0$. The isospin breaking coming from the
$D^{(\ast)\,\pm}-D^{(\ast)\,0}$ mass differences does introduce a
sizable $I=1$ component in the wave function of the
$X(3872)$~\cite{Ortega:2012rs}, but  we have checked that it does
  not alter the conclusions reached in this work.

We adopt the coupled-channels formalism described already in
Ref.~\cite{Ortega:2012rs} and decompose the hadronic state as
\begin{equation} \label{ec:funonda}
 | \Psi \rangle = \sum_\alpha c_\alpha | \psi_\alpha \rangle
 + \sum_\beta \chi_\beta(P) |\phi_A \phi_B \beta \rangle,
\end{equation}
where $|\psi_\alpha\rangle$ are $c\bar c$ eigenstates of the two body
Hamiltonian, 
$\phi_{M}$ are $q\bar q$  eigenstates describing 
the $A$ and $B$ mesons, 
$|\phi_A \phi_B \beta \rangle$ is the two meson state with $\beta$ quantum
numbers coupled to total $J^{PC}$ quantum numbers
and $\chi_\beta(P)$ is the relative wave 
function between the two mesons in the molecule. 

In this formalism, in addition to the direct meson-meson interaction
due to the exchange of pseudo-Goldstone bosons at $q\bar q$ level
described by the aforementioned CQM~\cite{Vijande:2004he}, with
parameters updated at Ref.~\cite{Segovia:2008zz} for the heavy quark
sectors, two- and four-quark configurations are coupled using the
$^{3}P_{0}$ model~\cite{LeYaouanc:1972vsx,LeYaouanc:1973ldf}, the same
transition mechanism that, within our approach, allows us to compute
open-flavor meson strong decays.
This model assumes that the transition operator
is 
\begin{eqnarray}
T&=&-3\sqrt{2}\gamma'\sum_\mu \int d^3 p d^3p' \,\delta^{(3)}(p+p')\times\nonumber\\&\times&
\left[ \mathcal Y_1\left(\frac{p-p'}{2}\right) b_\mu^\dagger(p)
d_\nu^\dagger(p') \right]^{C=1,I=0,S=1,J=0},
\label{TBon}
\end{eqnarray}
where $\mu$ ($\nu=\bar \mu$) are the quark (antiquark) quantum numbers and
$\gamma'=2^{5/2} \pi^{1/2}\gamma$ with $\gamma= \frac{g}{2m}$ is a dimensionless constant 
that gives the strength of 
the $q\bar q$ pair creation from the vacuum. 
From this operator we define the transition 
potential $h_{\beta \alpha}(P)$ within the $^{3}P_{0}$ model 
as~\cite{Kalashnikova:2005ui} 
\begin{equation}
\langle \phi_{A} \phi_{B} \beta | T | \psi_\alpha \rangle =
P \, h_{\beta \alpha}(P) \,\delta^{(3)}(\vec P_{\rm cm}).
\label{eq:Vab}
\end{equation}

Using the latter coupling mechanism, the coupled-channels system can be expressed as a Schr\"odinger-type equation,

\begin{equation}
\begin{split}
\sum_{\beta} \int \big( H_{\beta'\beta}(P',P) + &
V^{\rm eff}_{\beta'\beta}(P',P) \big) \times \\
&
\times \chi_{\beta}(P) {P}^2 dP = E \chi_{\beta'}(P'),
\label{ec:Ec1}
\end{split}
\end{equation}
where $H_{\beta'\beta}$ is the Resonating Group Method~(RGM) Hamiltonian for the two-meson states obtained from
the $q\bar{q}$ interaction. The effective potential $V^{\rm eff}_{\beta'\beta}$ encodes the coupling with the $c\bar c$ bare spectrum, and can be written as

\begin{equation}
V^{\rm eff}_{\beta'\beta}(P',P;E)=\sum_{\alpha}\frac{h_{\beta'\alpha}(P')
h_{\alpha\beta}(P)}{E-M_{\alpha}},
\end {equation}
where $M_\alpha$ are the masses of the bare $c\bar{c}$ mesons.

\begin{figure*}[htbp]
\begin{center}
\epsfig{figure=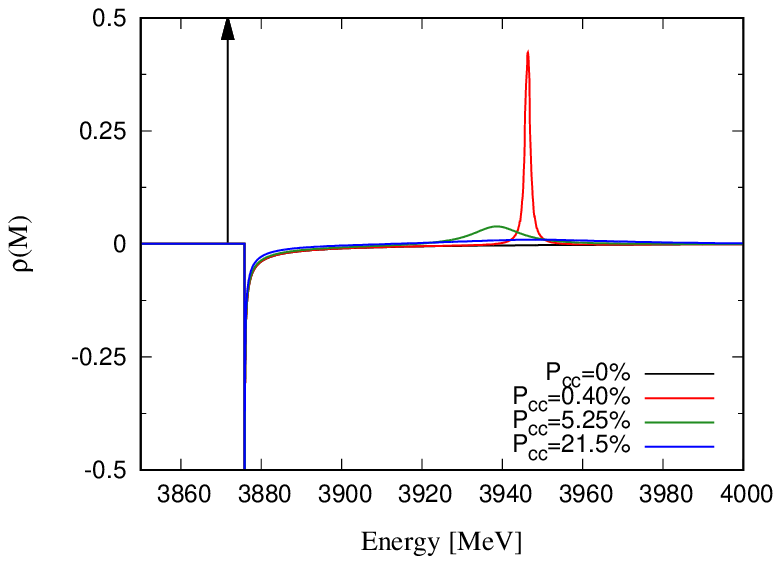,width=0.45\linewidth} 
\epsfig{figure=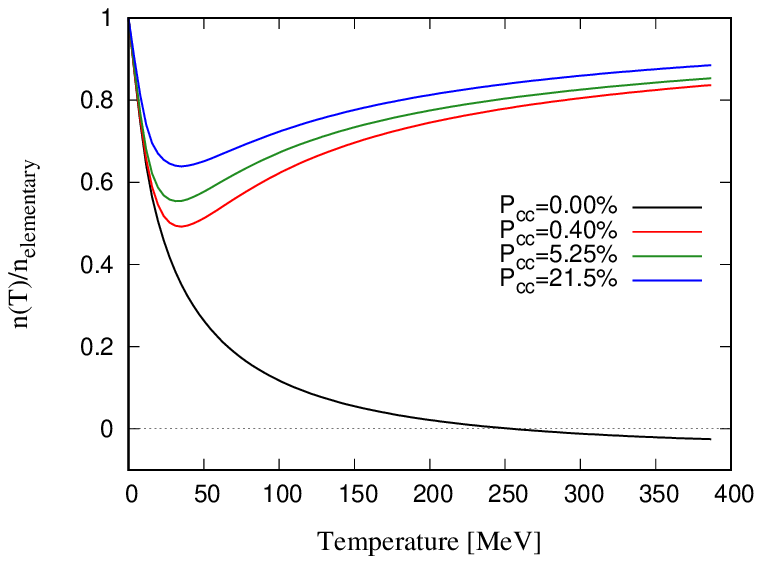,width=0.45\linewidth} 
\end{center}
\caption{Color online: Left panel: Total Level density $\rho(M)$
  (Eq.~\eqref{ec:stateden}) of the $D \bar D^*$ in the $J^{PC}=1^{++}$
  channel as a function of the mass. The arrow indicates the
  contribution of the $X(3872)$ bound state, which is a Dirac delta
  $\delta(m-m_X)$. Right panel: Occupation number $n(T)$ of the $D
  \bar D^*$ in the $J^{PC}=1^{++}$ channel, as a function of the
  temperature T (in MeV), with respect to the contribution
  of the $X(3872)$ assuming it is an elementary particle and no
  continuum contribution (Eq.~\eqref{ec:Nocup1}).}
\label{fig:ncluster}
\end{figure*}

In the cluster quark model picture the interaction between quarks
contains a tensor force due to pion exchange. Besides, the effective potential 
$V^{\rm eff}_{\beta'\beta}$ mixes different partial waves. Therefore, the S-matrix 
couples S and D waves, 
\begin{eqnarray}
S^{J1} &=&\left(
\begin{array}{cc}
\cos \epsilon_j & -\sin \epsilon_j \\ \sin
\epsilon_j & \cos \epsilon_j 
\end{array}
\right) \left(
\begin{array}{cc}
e^{2 {\rm i}
\delta^{1j}_{j-1}} & 0 \\ 0 & e^{2 {\rm i} \delta_{j+1}^{1j}} 
\end{array}
\right) \nonumber \\
&\times& 
\left( \begin{array}{cc}
\cos \epsilon_j & -\sin \epsilon_j \\ \sin
\epsilon_j & \cos \epsilon_j 
\end{array} 
\right) \, . 
\end{eqnarray}
From here we define the T-matrix
\begin{equation}
S^{JS}= 1 - 2 i k T^{JS} \, ,    
\end{equation}
The S and D eigen phase-shifts are shown in Fig.~\ref{fig:phases}
together with the result for the cumulative number. The outstanding
feature is the turnover of the function as soon as a slightly
non-vanishing $c \bar c$ content in the $X(3872)$ is included, unlike
the purely molecular picture. The steep rise in the phase shift
corresponds to a resonant state located at a mass $M\sim3945$MeV and
may be identified with the $X(3940)$ which in the purely molecular
picture would disappear as the $c\bar c$ spectrum would decouple from
the $D\bar D^\ast$ scattering.  Thus, the raise in the $1^{++}$
channel is not due to the $X(3872)$ but to the onset of the $X(3940)$
resonance.  The PDG values for X(3940)
  $M=3942(9)$MeV and $\Gamma=37^{+27}_{-17}$MeV~\cite{Patrignani:2016xqp} suggests indeed a
  non-vanishing mixing and $P_{\bar c c}= 5-25 \%$ for the X(3872).
 Moreover, we have checked that the S-wave phase-shift
  asymptotically approaches $\pi$ (due to the bound X(3940)-state of
  the purely confined channel) and hence $N(\infty)=\pi$ in agreement
  with the modified Levinson's theorem~\cite{Dashen:1976cf}.

\begin{table}
\begin{tabular}{c|c|c|c|c|c}
\hline $ \gamma(^3P_0) $ & $\mathcal{P}_{c\bar c}$ [\%] & $\alpha_0$
       [fm] & $r_0$ [fm] & $M$ [MeV] & $\Gamma$ [MeV]\\ \hline 0.00 &
       0.00 & 3.14 & 1.21 & 3947.43 & 0.00 \\ 0.05 & 0.40 & 3.14 &
       1.20 & 3946.29 & 1.38 \\ 0.10 & 1.82 & 3.11 & 1.17 & 3943.06 &
       5.88 \\ 0.16 & 5.25 & 3.05 & 1.10 & 3938.56 & 15.18 \\ 0.20 &
       14.25 & 2.88 & 0.85 & 3937.09 & 37.93 \\ 0.23 & 21.50 & 2.73 &
       0.63 & 3947.05 & 56.03 \\
\end{tabular}
\caption{$X(3872)$ $c\bar c$ probability, scattering length and
  effective range for the S-wave as a function of the dimensionless
  constant $\gamma$ of the $^3P_0$ transition operator. The mass of
  the $D\bar D^\ast$ bound state $X(3872)$ is fixed at $3871.7$ MeV.
  The mass and width of the $X(3940)$ resonance is also shown (PDG
  values are~\cite{Patrignani:2016xqp} $M=3942(9)$MeV and
  $\Gamma=37^{+27}_{-17}$MeV.)}
\label{tab:effective}
\end{table}

In Ref.~\cite{Ortega:2009hj} the $^3P_0$-model $\gamma$ parameter of
Eq.~\eqref{TBon} was constrained via strong decays in the charmonium
spectrum. However, in the present study we analyze the effect of
adiabatically connecting the $c\bar c$ spectrum and the $D\bar
D^\ast$, so we will vary $\gamma$ from zero to the value used in
Ref.~\cite{Ortega:2009hj}, maintaining the mass of the bound state
fixed at the experimental $3871.7$ MeV by consequently adapting the
strength of the direct meson-meson interaction. Besides this
re-scaling we take exactly the parameters of
Ref.~\cite{Ortega:2009hj}. The $X(3940)$ and the S-wave effective
range expansion parameters, are given in Table~\ref{tab:effective} for
different $\gamma$ values, where for the coupled-channels version of
Eq.~(\ref{eq:ERE}) we follow Ref.~\cite{PavonValderrama:2005ku}
adapted to the present situation. These values should be compared with
the lattice results \cite{Prelovsek:2013cra} for $m_\pi=266 {\rm MeV}$
of $\alpha_0 = 1.7 (4)$fm and $r_0 = 0.5 (1)$fm extracted from finite
volume calculations, bearing in mind that they found a binding energy
of $−11\pm7$ MeV below the $D^0\bar D^{0\,\ast}$ threshold.

\section{Finite temperature and X(3872) production}

Finally, we turn now to the consequences for finite temperature
calculations.  The level density and the corresponding occupation
number (relative to the elementary one) are shown in
Fig.~\ref{fig:ncluster} as functions of the invariant mass (left) and
the temperature (right). As we see that the cancellation between the
bound state and the continuum only happens for zero $c\bar c$
probability content, when the $c\bar c$ spectrum is decoupled from the
$D\bar D^\ast$ scattering. However, note that the non-vanishing
occupation number is merely due to the resonant reaction $D \bar D^*
\to X(3940) \to D \bar D^* $. This is exactly the same feature
observed in $\pi\pi$ scattering in the $1^{--}$ channel to to the $\pi
\pi \to \rho \to \pi \pi$ resonant
reaction~\cite{Venugopalan:1992hy,GomezNicola:2012uc,Broniowski:2015oha}.

Of course, one may wonder what is the range of applicability of the
present calculation, particularly as a function of the temperature. At
higher temperatures effects of hadron dissociation sets in,
accompanied by the explicit emergence of the quarks and gluons degrees
of freedom. The hadronic state representation would then, presumably,
break down. This is supported by recent lattice calculations, when
combinations of higher order fluctuations are computed
\cite{Karsch:2017mvg} and found to vanish for hadrons (in the
Boltzmann approximation) but not for quarks, and is found to be
non-vanishing for $T > 154$ MeV. Our Fig.~\ref{fig:ncluster} vividly
shows that the effect is quite visible before hadron dissociation, and
should thus be relevant in the study of production and absorption of
$X(3872)$ in a hot medium such as the one generated in heavy
  ion collisions~\cite{florkowski2010phenomenology}.

\begin{figure}[htbp]
\begin{center}
\epsfig{figure=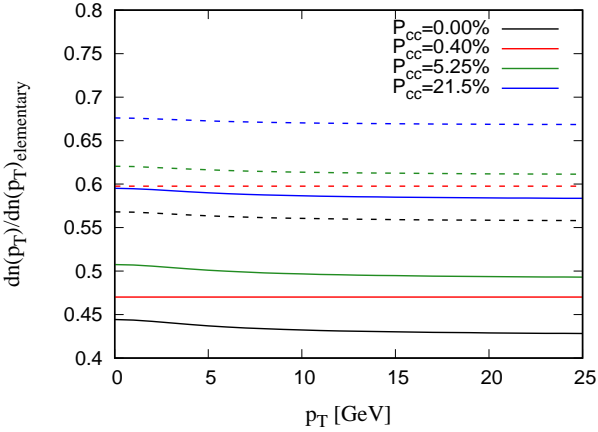,width=0.99\linewidth} 
\end{center}
\caption{Color online: Relative $p_T$ distribution (see Eq.~(\ref{eq:pT-n}))
of the $J^{PC}=1^{++}$-channel with a binning of $\Delta m= 2 B_X$MeV (dashed) 
$\Delta m= 5 B_X$ MeV (solid) for different $P_{\bar c c}$ content.
}
\label{fig:pT}
\end{figure}

The real experiments in pp-collisions uses a finite binning step
$\Delta m = 3$MeV~\cite{Chatrchyan:2013cld} and $\Delta m =
1.5$MeV~\cite{Aaboud:2016vzw}. We note that this is 10-15 times {\it
  much larger} than the binding energy of $B_X=0.01(18)$MeV quoted by
the PDG~\cite{Patrignani:2016xqp}. Therefore, any signal contains a
contamination of continuum and bound states in the $1^{++}$ channel
and it is foreseeable that future experiments in heavy ion collisions
will implement a similar $\Delta m$.

Actually, the $p_T$ distribution at mid-rapidity of a fireball at
rest~\cite{Schnedermann:1993ws} stemming from an invariant mass
distribution $\rho(m)$ binned with step $\Delta m$ in the notation of 
Eq.~(\ref{ec:Nocup1}) is given by 
\begin{eqnarray}
\frac{d \bar n (p_T)}{d m_T^2} = \int_{\Delta m} \rho(m) dm 
\sum_{n=1}^\infty  \frac{g m_T }{(2\pi)^2}  
\frac{(-\eta)^{n+1}}{n}
K_1 \left( \frac{n m_T}T \right) 
\label{eq:pT-n}
\end{eqnarray} 
where $m_T^2=p_T^2+m^2$ and the integral extends over $M_X \pm \Delta
m/2$.  The result of the ratio of the finite-$\Delta m$ binned to the
elementary $p_T$ distribution is shown in Fig.~\ref{fig:pT} for
$T=200$MeV. Neglecting isospin effects we have in the model
$B_X=4$MeV, so that we take $\Delta m = 2 B_X $ and $\Delta m = 5 B_X
$ to illustrate the situation. As we see, the effect is dramatic in
the strength which is reduced by almost $50\%$ and is saturated when
the binning is larger than $ \Delta m= 5 B_X$. We also see that the
$p_T$ dependence is not affected much in a wide range.  In a future
publication we will provide a more comprehensive analysis including
current freeze-out models, such as blast-wave or Hubble-like expansion
patterns which might realistically be tested with future heavy ion
X(3872) production experiments. This would require, in particular, a
fine tuning of parameters of Ref.~\cite{Ortega:2009hj} to account for
the most recent PDG figures~\cite{Patrignani:2016xqp}.

\section{Conclusions}

The production and absorption of $X(3872)$ in high energy heavy ion
collisions~\cite{Torres:2014fxa} or the time evolution of the
$X(3872)$ abundance in a hot hadron gas~\cite{Abreu:2016qci} has been
investigated recently in an attempt to pin down its structure from its
behavior in the Quark-Gluon Plasma (QGP). Abundances depend on the
nature of the state. These studies echo an opposite strategy with
similar studies of $J/\Psi$ where the melting of this very well known
state is used to diagnose the QGP. Our calculation shows that a
possible signal for $X(3872)$ abundance might in fact be erroneously
confused with the $X(3940)$ as a non-vanishing occupation number of
the $D \bar D^*$ spectrum in the $1^{++}$ channel at temperatures
above the crossover to the QGP phase. Below this temperature, the
$X(3872)$ does not count and should not be included in the Hadron
Resonance Gas. The Dashen-Kane effect extends also to $X(3872)$
  production and detection in heavy ions collisions and more generally
  in any production process where the experimental resolution exceeds
  the binding energy.

\section*{Acknowledgements}

We thank Wojcieh Broniowski and Alessandro Pilloni for helpful
remarks. This work has been partially funded by the Spanish Ministerio
de Economia y Competitividad and European FEDER funds (Grant
No. FIS2014-59386-P, FPA2016-77177-C2- 2-P), the Agencia de Innovacion
y Desarrollo de Andalucia (Grant No.  FQM225), and by Junta de
Castilla y Le\'on and European Regional Development Funds (ERDF) under
Contract no. SA041U16.

\bibliographystyle{h-elsevier}
\bibliography{Xcount}

\end{document}